\documentclass[amsmath,amssymb,floatfix,prb,twocolumn]{revtex4}

\usepackage{enumerate}
\usepackage{graphicx}
\usepackage{txfonts}
\usepackage{natbib}

\newcommand{\boldnabla}{\mbox{\boldmath$\nabla$}}

\begin{document}

\title{The onset of Impulsive Bursty reconnection at a two-dimensional current layer}
\author{J. Fuentes-Fern\'andez}
\email{jorge@mcs.st-andrews.ac.uk}
\author{C.~E. Parnell}
\author{E.~R. Priest}
\affiliation{Department of Mathematics and Statistics, University of St Andrews, North Haugh, St Andrews, KY16 9SS, United Kingdom} 

\begin{abstract}
The sudden reconnection of a non-force free 2D current layer, embedded in a low-beta plasma, triggered by the onset of an anomalous resistivity, is studied in detail. The resulting behaviour consists of two main phases. Firstly, a {\it transient reconnection} phase, in which the current in the layer is rapidly dispersed and some flux is reconnected. This dispersal of current launches a family of small amplitude magnetic and plasma perturbations, which propagate away from the null at the local fast and slow magnetosonic speeds. The vast majority of the magnetic energy released in this phase goes into internal energy of the plasma, and only a tiny amount is converted into kinetic energy. In the wake of the outwards propagating pulses, an imbalance of Lorentz and pressure forces creates a stagnation flow which drives a regime of {\it impulsive bursty reconnection}, in which fast reconnection is turned on and off in a turbulent manner as the current density exceeds and falls below a critical value. During this phase, the null current density is continuously built up above a certain critical level, then dissipated very rapidly, and built up again, in a stochastic manner. Interestingly, the magnetic energy converted during this quasi-steady phase is greater than that converted during the initial transient reconnection phase. Again essentially all the energy converted during this phase goes directly to internal energy. These results are of potential importance for solar flares and coronal heating, and set a conceptually important reference for future 3D studies.

\end{abstract}

\maketitle


\section{Introduction}

Many theoretical studies of magnetic reconnection have focused on steady-state reconnection since this is far simpler to analyse than non-steady reconnection and because many examples of reconnection, such as solar flares, are either steady or have a long-time quasi-steady component lasting for many Alfv\'en travel times. In the case of the solar flare, typically there are two intermingled components, namely, an ``impulsive'' phase lasting for 100-1000 secs with extremely rapid rises and falls and substructure over a second or less (e.g. \citet{hudson10}), and a ``gradual'' phase for which the soft X-ray and ${\rm H}\alpha$ emission slowly rises and falls (sometimes lasting a day or more). Other examples of non-steady reconnection include {\it flux-transfer events} at the Earth's magnetopause and the process of coronal heating \cite{walsh03,klimchuk06,hood10}, for which a prime candidate is the creation of {\it nanoflares} by reconnection in many small rapidly forming and dissipating current sheets, as described by Parker's braiding theory \cite{parker72} or its more effective update ({\it coronal tectonics} \cite{priest02a}) which takes account of the {\it magnetic carpet} \cite{schrijver97}.

However, it is important to understand the physical nature of two aspects of time-varying reconnection, namely, {\it transient reconnection} \cite{Longcope07,Fuentes12} caused by the sudden onset of an anomalous resistivity, and so-called {\it impulsive bursty reconnection} \cite{priest86b}, in which the reconnection continually either rapidly switches on and off or rapidly changes between slow and fast reconnection. This last regime can occur in a manner that is either quasi-periodic or random. The general observational consequences of these time-dependent reconnection regimes on the Sun are either smoothly varying or impulsive energy release (depending on the quasi-periodic timescale of bursts), particle acceleration, and generation of waves that propagate away from energy release sites. There are several physical ways in which impulsive bursty reconnection could in principle occur, one of them being closely related to a trigger mechanism for transient reconnection, i.e. in response to a magnetic diffusivity that depends on current density and that is greatly enhanced to a turbulent value when the current exceeds a threshold for microinstabilities. It is this mechanism that we focus on in the present paper.

In practice, reconnection takes place in three dimensions at several types of magnetic structures \cite{schindler88,priest96a,longcope05b}, namely, 3D null points \cite{pontin04,pontin05a,alhachami10,pontin11,priest09a}, separators \cite{longcope96a,longcope01,haynes07,parnell10a} and QSLs \cite{priest95a,demoulin96a,demoulin97b,aulanier06a}. In preparation for studying transient and impulsive bursty reconnection at such structures we focus here on the simpler, but conceptually important, problem of 2D reconnection, since the insights gained here will be invaluable for future 3D studies.

At two-dimensional magnetic X-points the way magnetic energy is converted into other forms in a steady manner has been a subject of study for decades. On the Sun, the continuous slow photospheric motions of the magnetic footpoints feed energy to the coronal magnetic field and store it in the form of force-free fields and current density layers \cite{parker72,vanBallegooijen85}. Locally within these current layers, the properties of the plasma may trigger the onset of reconnection. For instance, current layers may undergo current induced microinstabilities creating an anomalous resistivity \cite{Galeev84,Raadu88,Yamada97}, which then permits the dissipation of the current via reconnection. In this process, part of the magnetic energy is transferred into kinetic and/or internal energy of the plasma (plus particle acceleration in kinematic models). This is likely to be the case in solar flares \cite{Antiochos82,Barta11}, and is a possible mechanism for coronal heating via myriads of small-scale nano-flares \cite{Galsgaard96}. The onset of these instabilities, as well as the energy partitioning, are still not well understood.

Recently, the source and nature of energy conversion associated with 2D spontaneous reconnection at magnetic X-points have been considered by \citet{Longcope07}. They make an analytical study of the fast magnetosonic waves launched by reconnection in a current sheet after a sudden increase in the resistivity, neglecting the effects of plasma pressure. By conducting a one-dimensional analysis in which they investigate the leading order $m=0$ term only, they find a propagating sheath of current that travels out from the null at the local Alfv\'en speed, converting the magnetic energy into kinetic energy as it moves through the volume. Current dissipation is then modelled by introducing a uniform diffusivity, $\eta$, which is large enough for their dynamics to be approximated by the linear resistive magnetohydrodynamic (MHD) equations. 

Their analytical solution has two distinct parts: the first is described by a diffusion equation and the second by a propagating wave equation. As soon as the resistivity is enhanced, diffusion expands the current rapidly (at a rate much quicker than any wave travel time), and this current expansion then slows down to the point where the speed of expansion couples to the local FMS mode. From that point on, a propagating fast wave expands outwards from the edge of the diffusion region, carrying most of the energy converted in the reconnection process.

Based on \citet{Longcope07}, \citet{Fuentes12} considered this same question numerically, starting from a non-force-free numerical equilibrium obtained by \citet{Fuentes11} through the ideal MHD evolution of a perturbed magnetic X-point embedded in a high-beta plasma. In this equilibrium, a thick current layer is formed at the location of the null, that extends along the four separatrices. The current density at the location of the null is slowly evolving in an asymptotic manner towards an infinite time singularity, while the central current layer becomes gradually thinner and shorter in time.

Starting from the quasi-static equilibrium described above, \citet{Fuentes12} studied the consequences of spontaneous reconnection \cite{Longcope94,Craig94b} after the sudden onset of an anomalous resistivity in a high plasma beta regime. The present study follows on from this, but instead considers a lower beta plasma case. In order to compare and understand the differences between the two experiments, we briefly summarise here the key characteristics and results from \citet{Fuentes12}:\\
1) The amount of magnetic energy available for the reconnection process, and the rate of conversion of this energy, depend on the value of the anomalous resistivity, in contrast to the steady-state reconnection models, such as some fast reconnection regimes \cite{Petschek64,Biskamp86,priest86b}, where the rate with which the reconnection is being driven is the same as the actual rate with which the flux is reconnected.\\
2) The resistivity rapidly diffuses the current, converting magnetic energy into both kinetic and internal energy, and expanding slightly the initial current layer. Most of the transferred energy during the reconnection is directly used to heat the plasma.\\
3) By the end of the diffusion phase, the speed of expansion of the current layer couples with the local magnetosonic speeds and various pulses are launched out of the diffusion region. These pulses transport both a plasma and a magnetic perturbation outwards from the null. Due to the high values of the plasma beta, the slow and fast magnetosonic speeds are very similar, and the overall expansion is nearly circular.\\
4) The forces that are responsible for these perturbations are such that in their wake there remains a reconnection velocity flow pattern (inwards above and below the null and outwards to the sides).\\

The numerical relaxation of magnetic fields and plasmas driven by physical viscous forces to form the current layer equilibrium field is complex and computationally demanding. In the past, several people have used a fictitious damping mechanism in magneto-frictional codes. In addition, numerical resolution is a tough problem that has to be faced when doing numerical simulations. In our particular case, the smaller the plasma pressure, the thinner the current structures will become, and hence, the better the resolution needs to be. There have been a number of studies of 2D current sheet formation and current singularities using AMR (adaptive mesh refinement) codes \cite{Friedel96,Grauer98}. Our approach here is to use the same code as \citet{Fuentes11} and \citet{Fuentes12}, with a high resolution homogeneous grid, and to focus on the nature of the subsequent reconnection process, rather than the formation of the singularity, which would require higher resolution. 

The paper is structured as follows. In Sec. \ref{sec:sec2}, we briefly describe the numerical code and state the equations governing the evolution of the system. In Sec. \ref{sec:sec3}, we present the initial non-resistive evolution of the field, and the formation of a thin current layer at the location of the null, which serves as a starting point for the reconnection experiments, described in in Sec. \ref{sec:sec4}. Finally, we end with a summary and some discussion in Sec. \ref{sec:sec5}.


\section{Numerical scheme and setup} \label{sec:sec2}

\begin{figure*}[t]
  \centering
  \includegraphics[scale=0.62]{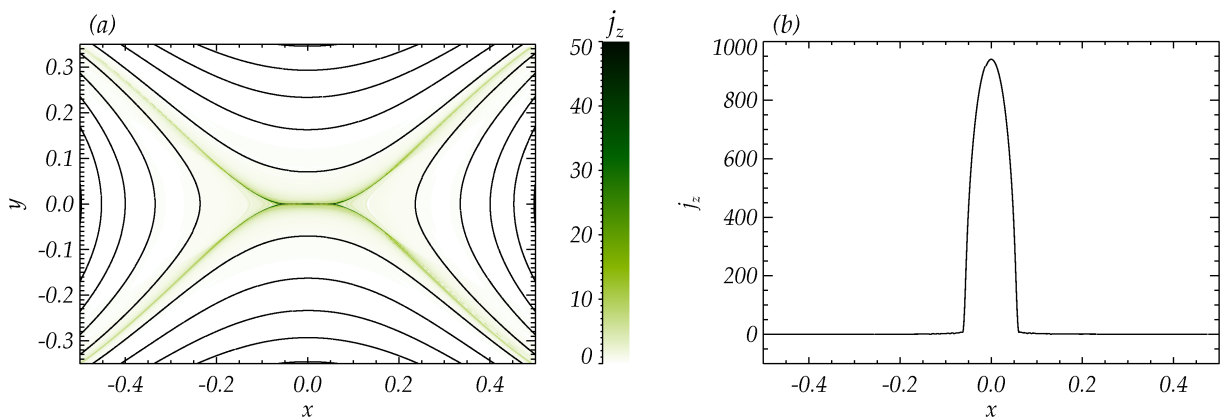}
  \caption{Plots of (a) the current density map and magnetic field lines, and (b) a horizontal cut of current density along the $x$-axis of the central current layer, in the non-force-free equilibrium state reached after the non-resistive relaxation. In (a) the current levels have been cut-off at $j_z=50$ in order to see the extension along the four separatrices.}
  \label{fig:current1}
\end{figure*}

For the numerical experiments studied in this paper, we have used Lare2D \cite{Arber01}, a staggered Lagrangian-remap code with user controlled viscosity and resistivity, that solves the full MHD equations. The staggered grid is used to build conservation laws and maintains $\boldnabla\cdot{\bf B}=0$ to machine precision, by using the Evans and Hawley's constrained transport \cite{Evans88} method for the magnetic flux.

The numerical code uses the normalised MHD equations where the normalised magnetic field, density and lengths,
\begin{eqnarray*}
x=L\hat{x}\;,\;\;\;y=L\hat{y}\;,\;\;\;{\bf B}=B_n\hat{\bf B}\;,\;\;\;\rho=\rho_n\hat{\rho}\;,
\end{eqnarray*}
imply that the normalising constants for pressure, internal energy, current density and plasma velocity are,
\begin{eqnarray*}
p_n=\frac{B_n^2}{\mu}\;,\;\;\;\epsilon_n=\frac{B_n^2}{\mu\rho_n}\;,\;\;\;j_n=\frac{B_n}{\mu L}\;\;\;{\rm and}\;\;\;v_n=\frac{B_n}{\sqrt{\mu \rho_n}}\;.
\end{eqnarray*}
The subscripts $n$ indicate the normalising constants, and the {\it hat} quantities are the dimensionless variables used in the code. The expression for the plasma beta can be obtained from this normalization as
\begin{eqnarray*}
\beta=\frac{2\hat{p}}{\hat{B}^2}\;.
\end{eqnarray*}
In this paper, we will work with the normalised quantities, but the {\it hat} is removed from the equations for simplicity.

The (normalised) equations governing our MHD processes are,
\begin{eqnarray}
\frac{\partial \rho}{\partial t}+\boldnabla\cdot(\rho{\bf v}) &=& 0\;,\label{n_mass}\\
\rho\frac{\partial{\bf v}}{\partial t}+\rho({\bf v}\cdot\boldnabla){\bf v} &=& -\boldnabla p + (\boldnabla\times{\bf B})\times{\bf B} + {\bf F}_{\nu}\;,\label{n_motion}\\
\frac{\partial p}{\partial t}+{\bf v}\cdot\boldnabla p &=& -\gamma p \boldnabla\cdot{\bf v}+H_{\nu}+\frac{j^2}{\sigma}\;,\label{n_energy}\\
\frac{\partial{\bf B}}{\partial t} &=& \boldnabla\times({\bf v}\times{\bf B})+\eta\nabla^2{\bf B}\;,\label{n_induction}
\end{eqnarray}
where $\eta$ is the magnetic diffusivity (which, in the normalised equations, equals the resistivity), ${\bf F}_{\nu}$ and $H_{\nu}$ are the terms for the viscous force and viscous heating, and $j^2/\sigma$ is the ohmic dissipation. The internal energy, $\epsilon$, is given by the ideal gas law, $p=\rho\epsilon(\gamma-1)$, with $\gamma=5/3$.

Magnetic field lines are line-tied at the four boundaries and all components of the velocity are set to zero on the boundaries. The other quantities have their derivatives perpendicular to each of the boundaries set to zero. This means that there are no losses of mass or energy through the boundaries, so these two quantities are conserved in the experiment, to within numerical error.

Finally, the time unit used in our numerical results is the fast mode crossing time, $t_F$, i.e. the time for a fast magnetosonic wave to travel from the null ($y=0$) to the top ($y=0.35$) or bottom boundary. This time is calculated analytically as
\begin{equation}
t_F=\int_{y=0}^{y=0.35}\!\frac{{\rm d}y}{c_F(y)}\;,
\end{equation}
where $c_F(y)=\sqrt{v_A^2+c_s^2}$ is the local fast magnetosonic speed.


\section{Non-resistive viscous relaxation} \label{sec:sec3}

By setting $\eta=0$ (and $\sigma\to\infty$) in Equations (\ref{n_mass}) to (\ref{n_induction}) above, we can eliminate the last terms of Equations (\ref{n_energy}) and (\ref{n_induction}), obtaining the non-resistive MHD equations with which no reconnection is permitted, leaving aside numerical effects.

The initial state for the non-resistive experiment is a hyperbolic X-point, $A_z=(x^2-y^2)/2$, squashed in the vertical $y$-direction by a given amount $(1-h)$ times the height of the original system, such that the flux function of the initial state is given by
\begin{equation}
A_z(x,y,0)=\frac{1}{2}\left(x^2-\frac{y^2}{h^2}\right)\;.
\end{equation}
The squashing creates a uniform current density whose only non-zero component is
\begin{equation}
j_z(x,y,0)=\frac{1}{h^2}-1\;.
\end{equation}
The squashing is characterised by the height of the box, normalised to the original height, $h$. The initial plasma pressure $p_0$, is set to a constant everywhere, such that the initial system is not in equilibrium. There are no initial flows in the domain.

\begin{figure}[t]
  \centering
  \includegraphics[scale=0.82]{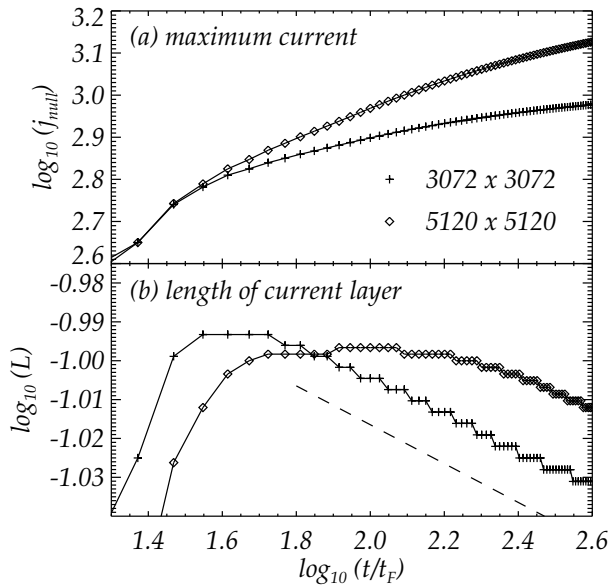}
  \caption{Time evolution of (a) the null current density and (b) the length of the central current (full width half maximum), for two experiments with different numerical resolution. In (b), the dashed line is a power law with slope $-0.05$.}
  \label{fig:currentsheet}
\end{figure}

For the experiment studied in this paper, the height of the box is $h=0.7$, the initial background values of the current density and the plasma pressure are $j_0=1.04$ and $p_0=0.05$, and the viscosity is isotropic and has a value of $\nu=0.0001$. The resolution of the numerical grid is $3072\times 3072$, which uniformly covers the domain extending from $-0.5$ to $0.5$ in the $x$-direction and $-0.35$ to $0.35$ in the $y$-direction.

The dynamical evolution of the system, driven by the viscous forces, concentrates the current density in a thin and elongated current layer at the location of the null point, as seen in Fig. \ref{fig:current1}a. The vast majority of the total current density in the final equilibrium state is confined to the central layer, whose width is less than the size of a numerical grid-cell. Hence, for a small background plasma pressure, we are no longer able to resolve the current structures, in contrast with the larger plasma pressure experiments in \citet{Fuentes11}. In addition to the main current layer, a significant current density extends along the separatrices, as already mentioned in previous non-force-free studies \cite{Rastatter94,Craig05,Pontin05,Fuentes11}. But this time, in low value plasma pressure case, a different scenario is found, in which the length of the central current layer is well defined (Fig. \ref{fig:current1}b), and the current along the separatrices is only 0.1\% of the value of the peak current (Fig. \ref{fig:current1}b).

By the end of the viscous relaxation, the system has entered an asymptotic regime and is heading towards an infinite-time singularity owing to tiny residual forces which very slowly build up the current at the central layer (for a detailed discussion of this behaviour, see \citet{Fuentes11}). In Fig. \ref{fig:currentsheet}a we show the time evolution of the peak current density (i.e. the current density at the null), for two experiments with different numerical resolution, namely $3072^2$ and $5120^2$. The same qualitative behaviour is observed, although the growth rate of the singularity is larger for the highest resolution. This is due to the fact that the central current layer is not well resolved in width. Although both current layers contain the same total current, the higher resolution experiment has a thinner current sheet, and hence, a larger peak current. Fig. \ref{fig:currentsheet}a shows the infinite-time singularity being formed, but the value of the growth rate is not physical, due to the resolution problem. This problem did not arise in experiments with higher background plasma pressure values \cite{Fuentes11}, which allow the current structures to be thicker (thus, resolvable), but shorter.

\begin{figure*}[t]
  \centering
  \includegraphics[scale=0.62]{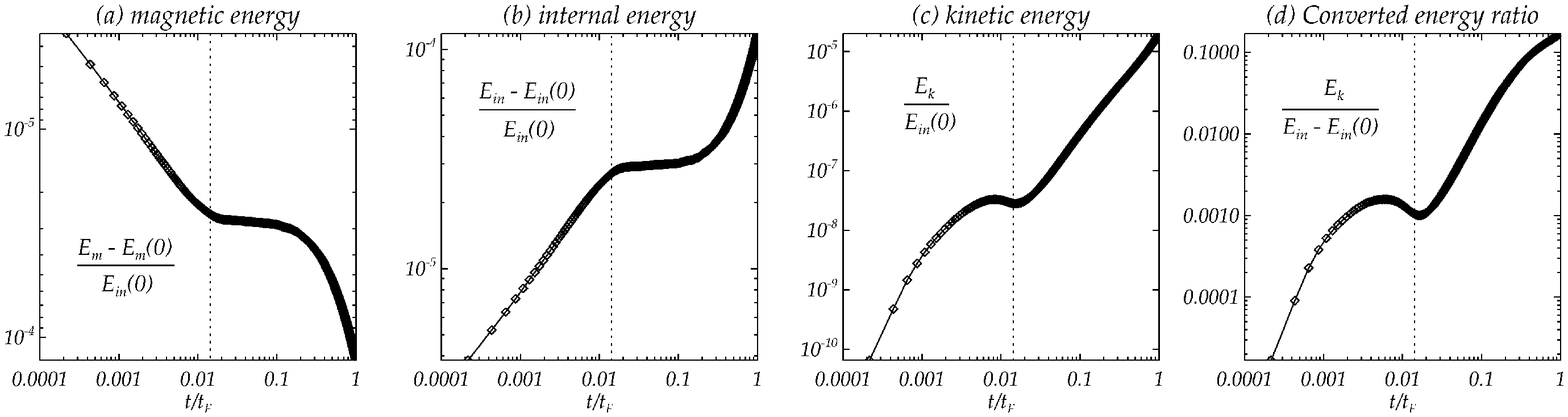}
  \caption{Time evolution of the different energies of the system, after the reconnection has begun, with logarithmic axes. Showing, the change in a) magnetic, b) internal and c) kinetic energy. In d) we show the ratio of the change in kinetic energy to the change in internal energy. Note, that the initial kinetic energy is zero. The initial values of magnetic energy and internal energy are $0.08$ and $0.06$, respectively. The vertical dotted line indicates the time at which the diffusive regime finishes.}
  \label{fig:energies1}
  \vspace{0.5cm}
  \includegraphics[scale=0.62]{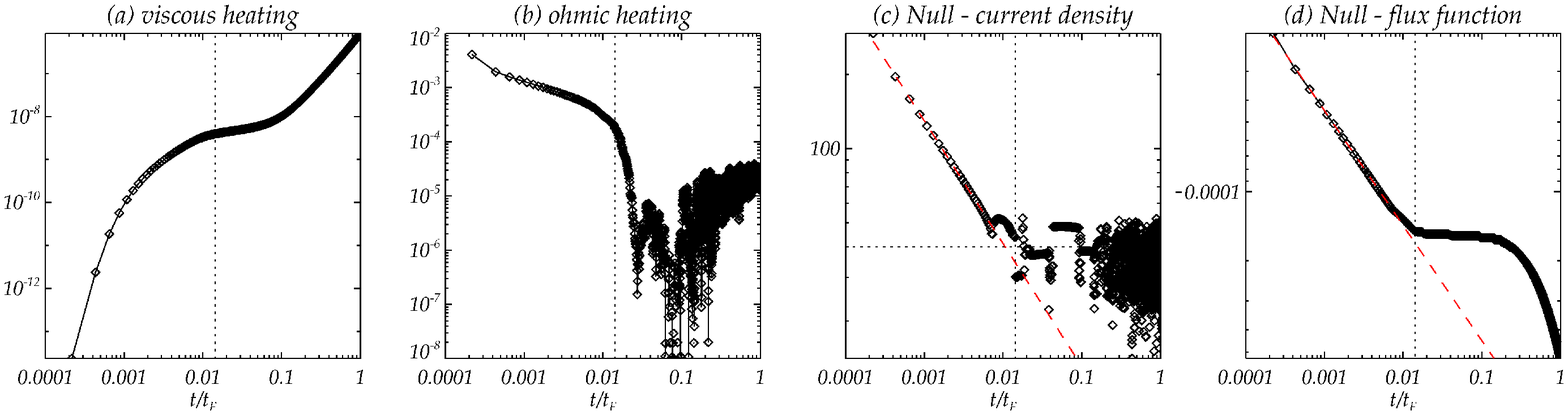}
  \caption{Logarithmic time evolution of a) the viscous heating, b) the ohmic heating, c) the current density at the null and d) the flux function at the null. The vertical dotted line indicates the time where the null current density goes below the level of  $j_{crit}=40$ (the dotted horizontal lines in c), i.e. where the diffusive regime finishes. For the null current density and flux function (c and d), the diffusive behaviour is modelled with a function of the form $c\,(t/t_F)^{-d}$, with $d=0.5$ in both cases.}
  \label{fig:energies2}
\end{figure*}

In Fig. \ref{fig:currentsheet}b, we show the time evolution of the length of the current sheet, calculated as the full width at half maximum, for the same two resolutions as in Fig. \ref{fig:currentsheet}a. The overall behaviour is the same for both experiments. During the first part of the evolution (i.e. the viscous relaxation), the central current layer gets more elongated as time elapses. However, when the viscous forces stop acting, and the field enters the asymptotic regime, then the length of the current layer starts decreasing. This is in agreement with the results from \citet{Fuentes11}, where the initial current layers were well resolved. The length of the central current layers following a negative power law with slope $-0.05$ (Fig. \ref{fig:currentsheet}b).

In contrast with the force-free scenarios studied analytically in the past \cite{priest86a,Klapper97}, where the magnetic field is believed to collapse towards a long and thin current sheet, here, the effect of the plasma pressure is to drive an infinite-time singularity by concentrating the current at the location of the null, causing the current layer to change its shape, hence reducing its length gradually.


\section{Reconnection experiment} \label{sec:sec4}

The non-force-free quasi-equilibrium state found above is the starting point for our reconnection experiment. The equations governing the evolution of the field are now the full MHD equation set given by Equations (\ref{n_mass}) to (\ref{n_induction}). The resistivity is anomalous. It vanishes below a critical value of the current density, $j_{crit}$, and is constant above it,
\begin{equation}
  \eta =
  \left\{
  \begin{array}{ll}
    0  & \mbox{if } j_z < j_{crit}\;, \\
    \eta_0 & \mbox{if } j_z \ge j_{crit}\;.
  \end{array}
  \right.
\end{equation}
This approach allows diffusion only about the null, where the current density is large. For the experiment analysed here, we have chosen $j_{crit}=40$ and $\eta_0=0.001$. The estimated numerical diffusivity, due to the finite grid resolution, is $0.0001$. In the absence of the enhanced diffusivity, the current layer would keep evolving slowly in time becoming shorter and thinner, with its peak current increasing until the onset of the numerical diffusion through which reconnection would occur resulting in a decrease in peak current. No such reduction in the peak current is evident during the formation of the initial equilibrium, thus we are confident that the experiment has not reached the numerical diffusion limit.


\subsection{Energetics}

In Fig. \ref{fig:energies1}, we show the changes in magnetic, internal and kinetic energies, normalised to the value of the initial internal energy, and the ratio of the change in kinetic energy to the change in internal energy. Note, that the kinetic energy is zero in the initial equilibrium state. In Fig. \ref{fig:energies2}, we show the time evolution of the viscous and ohmic heating, and the current density and flux function at the location of the null. In all these plots, the time interval shown extends from the moment the reconnection starts until the instant that a fast magnetosonic pulse, launched from the null at $t/t_F=0$, arrives at the upper boundary ($t/t_F=1$).

We may divide the evolution of the system in two parts. The first extends from $t/t_F=0$ to when the null current density first decreases to the critical value $j_{crit}=40$ ($t/t_F\sim 0.014$), below which there is no diffusion. This first energetic phase is associated with the transient diffusion which eliminates the pronounced current density peak obtained in the previous ideal relaxation (this is studied in Sec. \ref{sec:transient}). Right after the transient reconnection phase, the system enters a quasi-steady state in which reconnection seems to keep occurring at the null (as indicated by the flux function in Fig. \ref{fig:energies2}d), but in a chaotic manner, as shown by the behaviour of the null current density and the ohmic heating (Fig. \ref{fig:energies2}c and b, respectively). This {\it impulsive bursty reconnection} (studied in Sec. \ref{sec:bursty}) is produced by a continuous build-up of current density at the null in an environment with an anomalous resistivity that switches off below a given critical value of the current density.

During the first reconnection phase, we can see the sudden decrease of magnetic energy (Fig. \ref{fig:energies1}a), due to transient magnetic reconnection occurring at the null, which is directly transferred into both internal (Fig. \ref{fig:energies1}b), and kinetic energy (Fig. \ref{fig:energies1}c). The relative changes of energy follow a power law distribution. These changes in energy are very small, due to the very narrow initial diffusion region, which only allows a small amount of magnetic flux to be reconnected within the transient diffusion phase. However, the subsequent impulsive bursty phase is shown to convert much more magnetic energy in the long term, both into internal and kinetic energy, and it does it faster than by following a simple power law. This impulsive bursty regime is not present in the high beta case \cite{Fuentes12}, as will be discussed in Sec. \ref{sec:bursty}.

The ratio of converted internal energy to kinetic energy (Fig. \ref{fig:energies1}d) in the transient phase is surprisingly small (about 0.01), and increases rapidly to a value of 0.1 by the end of the experiment. Therefore, most of the released magnetic energy directly heats the plasma, in contrast with past studies \cite{Longcope07} where plasma pressure effects were neglected.

In Fig. \ref{fig:energies2}a and b, we show the time evolution of the integrated viscous and ohmic heating, respectively. As indicated by the energy partitioning, the viscous damping term, caused by the dissipation of the kinetic energy, is much smaller than the ohmic heating coming directly from the magnetic field dissipation. The impulsive bursty phase shows a complex non-smooth pattern which will be evaluated later, in Sec. \ref{sec:bursty}.


\subsection{Transient phase} \label{sec:transient}

The onset of an anomalous resistivity produces a sudden reconnection which very quickly diffuses the current density below the level given by $j_{crit}$. During this transient phase, the current density and the flux function at the null point, decrease following a power law of the form $c\,(t/t_F)^{-0.5}$  (see figures \ref{fig:energies2}c and d, respectively).

We want to determine the dependence of this diffusion rate on the value of the resistivity. But first, we consider the diffusion of a 1D current sheet given by a magnetic field ${\bf B}=B(x,t)\,{\bf e}_z$, with
\begin{equation*}
B(x,0) = \begin{cases}
B_0, &{\rm if}\; x>0\,, \\
-B_0, &{\rm if}\; x<0\,.
\end{cases}
\end{equation*}
The generic solution to the diffusion equation, $\partial{\bf B}/\partial t=\eta\nabla^2{\bf B}$, for small values of $|x|$, is
\begin{equation*}
B(x,t)=B_0\,{\rm erf}\left(\frac{x}{\sqrt{4\eta t}}\right)\approx B_0\frac{x}{\sqrt{\pi\eta t}}\;.
\end{equation*}
Therefore, assuming ${\bf j}=j(x,t)\,{\bf e}_y$ and $B=-\partial A/\partial x$, we get
\begin{eqnarray}
j(0,t)&=&\frac{1}{\eta}\frac{\partial B}{\partial x}\sim \frac{1}{\sqrt{\eta t}}\;, \label{solj} \\
A(0,t)&=&-\int_{\sqrt{\pi\eta t}}^0\!B\,dx\sim \sqrt{\eta t}\;. \label{solA}
\end{eqnarray}

\begin{figure}[t]
  \centering
  \includegraphics[scale=0.51]{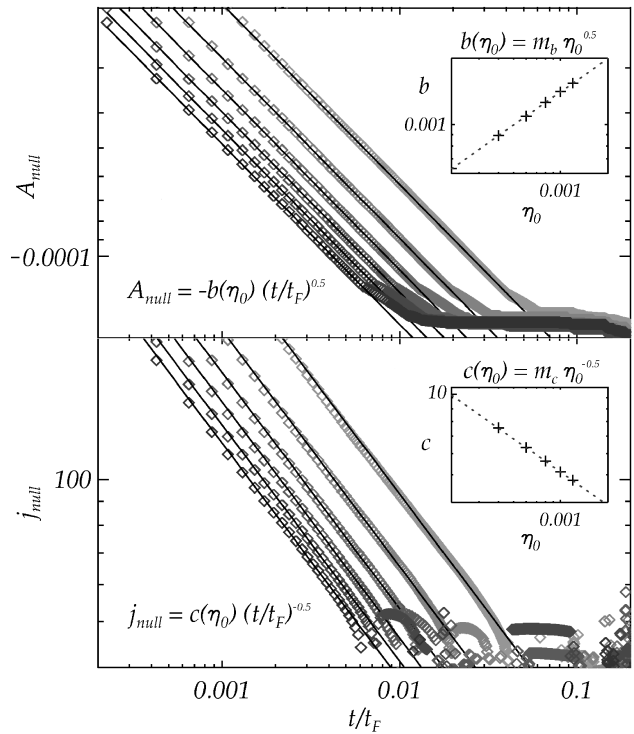}
  \caption{Logarithmic plots of the null flux function (top) and null current density (bottom), for the initial transient reconnection phase, for six experiments with different values of the resistivity, $\eta_0$, ranging from $0.0002$ to $0.0012$ (from left to right in the plots). All these are modelled with functions of the form $A_{null}=b(\eta_0)(t/t_F)^{0.5}$ and $j_{null}=c(\eta_0)(t/t_F)^{-0.5}$. Then, the functions $b(\eta_0)$ and $c(\eta_0)$ are plotted (top and bottom inserts, respectively), again using logarithmic axes, showing further power laws as a function of $\eta_0$.}
  \label{fig:dif_etas}
\end{figure}

Now, to determine the functionality of our numerical solution with resistivity, we run a series of experiments for four different values of $\eta_0$. In Fig. \ref{fig:dif_etas}, we show plots of the null flux function and the null current density during the transient reconnection phase, for 6 different experiments with $\eta_0$ ranging from $0.0002$ to $0.0012$. Each of these plots is modelled with a function of the form
\begin{eqnarray}
A_{null}&=&-b(\eta_0)\,(t/t_F)^{0.5}=-m_b\sqrt{\eta_0\,t/t_F}\;, \label{anull} \\
j_{null}&=&c(\eta_0)\,(t/t_F)^{-0.5}=\frac{m_c}{\sqrt{\eta_0\,t/t_F}}\;, \label{jnull}
\end{eqnarray}
in an equivalent manner as in Equations (\ref{solj}) and (\ref{solA}). Note, that the experiment is different from the analytical study in \citet{Longcope07}, in which they consider the solution of the current in an infinitesimally thin wire, as $j_{null}=C_0/(2\eta_0 t)$.

It is clear from Fig. \ref{fig:dif_etas} that the diffusion time depends on the value of the resistivity. This diffusion time may be calculated as the time that the current takes to reach a specific value of $j_{null}$, for all the different values of $\eta_0$. This calculation results in an expression of the form form $t_D=\kappa\eta_0^{-1}$, i.e. the diffusion time is inversely proportional to the value of the resistivity.

\begin{figure*}[t]
  \centering
  \includegraphics[scale=0.69]{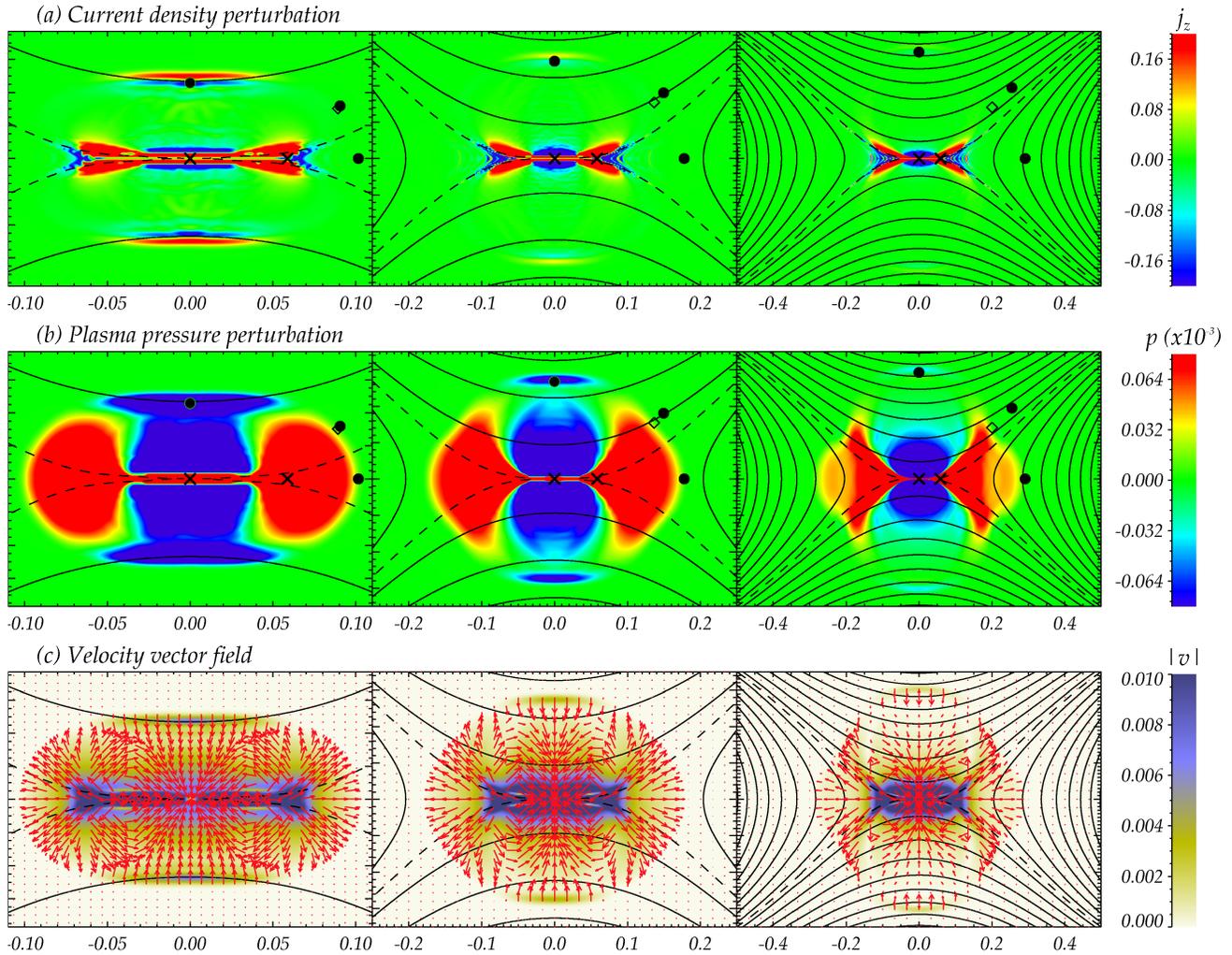}
  \caption{Contour plots of (a) current density perturbation, (b) plasma pressure perturbation and (c) velocity field (where red arrows indicate the direction of the velocity vector field at each point), with isotropic axes. In order to show the expanding motion of the pulses, we show three different times, namely $t=0.18t_F$ (left), $t=0.47t_F$ (center) and $t=0.82t_F$ (right). Black dots move radially at the local fast magnetosonic speed, from the null point and from the edge of the diffusion region (starting points are marked with an `x'). The diamond moves at the slow magnetosonic speed, radially from the edge of the diffusion region. Note, that the pressure bar values are to be multiplied by $10^{-3}$.}
  \label{fig:wavemap}
\end{figure*}

Combining now Equations (\ref{anull}) and (\ref{jnull}), we find
\begin{equation}
\frac{{\rm d}A_{null}}{{\rm d}t}=-\xi\,\eta_0 j_{null}=-0.17\,\eta_0 j_{null}\;.
\end{equation}
If during this transient phase, the current at the null had decreased purely due to the reconnection, then, $\xi$ should have the value of $1$. Instead, reconnection accounts for just $17\%$ of the change in current density at the null. The other $83\%$ is a result of the rapid and very localised dispersal of the peak current density. Under the effects of an anomalous resistivity, this leads to a decrease in the total reconnection occurring during the initial transient phase. 

On the other hand, the rate of the transient reconnection is completely independent of the critical current, $j_{crit}$. This means that the duration of the reconnection increases as $j_{crit}$ decreases. This has been established numerically by running experiments with different values of $j_{crit}$. The initial current density diffusion and transient reconnection occur indistinguishably for all cases, apart from their duration: the value of the critical current defines only the instant at which this reconnection phase finishes.

\begin{figure*}[t]
  \centering
  \includegraphics[scale=0.5]{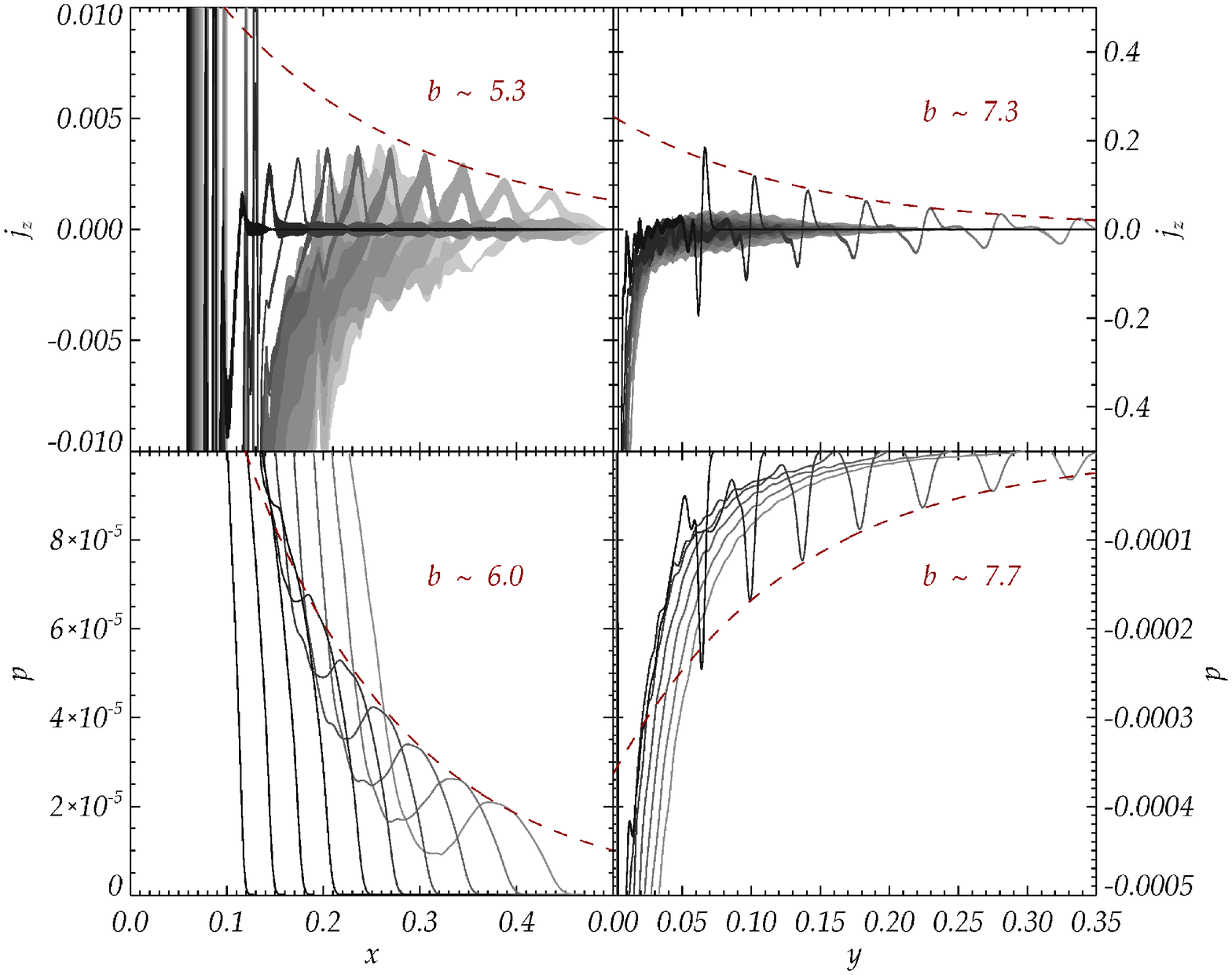}
  \caption{One-dimensional plots of the current density perturbation (top) and the plasma pressure perturbation (bottom) along the $x$-axis and the $y$-axis (left and right respectively). A range of different times are overplotted, in order to show the propagation of the perturbations both in the horizontal and in the vertical direction. In each case, the damping of the pulses is modelled with an envelope of the form $a\,{\rm e}^{-br}$ (red dashed lines), where $r$ can be either $x$ or $y$. The damping rates are shown in the figure.}
  \label{fig:1Dplots2}
\end{figure*}


\subsection{Wave initiation and propagation} \label{sec:wave}

Immediately after the short diffusive phase, various magnetosonic waves are launched in all directions. In Figures \ref{fig:wavemap}a, b and c, we show contour plots of perturbations in current density, plasma pressure, and magnitude of the velocity, for three different times during the propagation of the wave pulses, namely at $t=0.18t_F$, $t=0.47t_F$ and $t=0.82t_F$.

In the first time frame (left column in the plot), we see that a pair of planar current density pulses are launched from the top and the bottom of the diffusion region, each of these carrying a positive and a negative component of current. The plasma pressure map in the same frame shows a negative perturbation (deficit) which corresponds to the planar current density perturbation described before. In addition, we can see two semicircular positive pulses coming from the left and right edges of the diffusion region. The outwards propagation of these pulses can be clearly seen from the velocity map. Here, a negative pressure perturbation corresponds to an inwards velocity field, and a positive pressure perturbation corresponds to an outwards velocity field. Note the circular shape of the pulses coming from the two vertices.

In an attempt to track the propagation speed of these structures, we have added four symbols in the current density and plasma pressure maps. Black dots move radially at the local {\it fast magnetosonic speed}, one upwards from the null point (marked with an `x'), and two from the right edge of the diffusion region (also marked with an `x'), moving horizontally and diagonally (at an angle of $45^{\circ}$). The diamond moves at the local {\it slow magnetosonic speed}, from the right edge of the diffusion region, moving at an angle of $45^{\circ}$.

In the second time frame (center column in the plot) we see how the propagating pulses of current density have decayed in amplitude. The two semi-circular pulses, as can be seen from the pressure and velocity plots, are losing their radial symmetry, and instead, the horizontal motion seems to be faster than the diagonal one. Looking at the third time frame and comparing the position of the wave front with the black dots and the diamond, we can conclude that both the vertical and the horizontal propagations (that move across the magnetic field lines) are moving with the fast magnetosonic speed. In contrast, the diagonal propagations (that move roughly along the magnetic field lines) move with the slow magnetosonic speed, as one might expect for such a non-uniform magnetic field configuration.

The whole domain is squashed in the vertical direction, and the plasma pressure is greater above and below the diffusion region than it is to the left and right of it. Thus, the fast magnetosonic speed grows faster in the vertical than in the horizontal direction. Hence, despite the fact that the initial overall wave structure is elongated in the $x$-direction, its geometry changes such that, finally, it is more elongated in the vertical $y$-direction.

In order to evaluate the damping of the propagating pulses, in Fig. \ref {fig:1Dplots2}, we show horizontal and vertical cuts of both the current and the plasma pressure perturbations, for seven different times over the propagation of the wave, before it reaches any of the boundaries of the system. For all cases, we have fitted an envelope of the form $a\,{\rm e}^{-br}$, in accordance with the standard Fourier solution for a damping wave. The damping rates obtained are all very similar to each other and show a clear fading of the waves as they move throughout the domain. This fading may be due to two effects. First, the perturbations expand in area as they travel, so they decrease in amplitude. Second, the viscous terms act to slowly convert kinetic energy into internal energy of the system. This effect is not really significant as an overall contribution to the internal energy of the system (See Figures \ref{fig:energies2}a and b). Therefore, in contrast with the results from the experiment by \citet{Longcope07}, the propagating wave does not release a significant amount of energy away from the null. This is due to the fact that the initial amount of magnetic energy used in accelerating the plasma in the form of different magnetosonic waves is much smaller than the energy used in direct heating.


\subsection{Impulsive bursty reconnection} \label{sec:bursty}

\begin{figure}[t]
  \centering
  \includegraphics[scale=0.56]{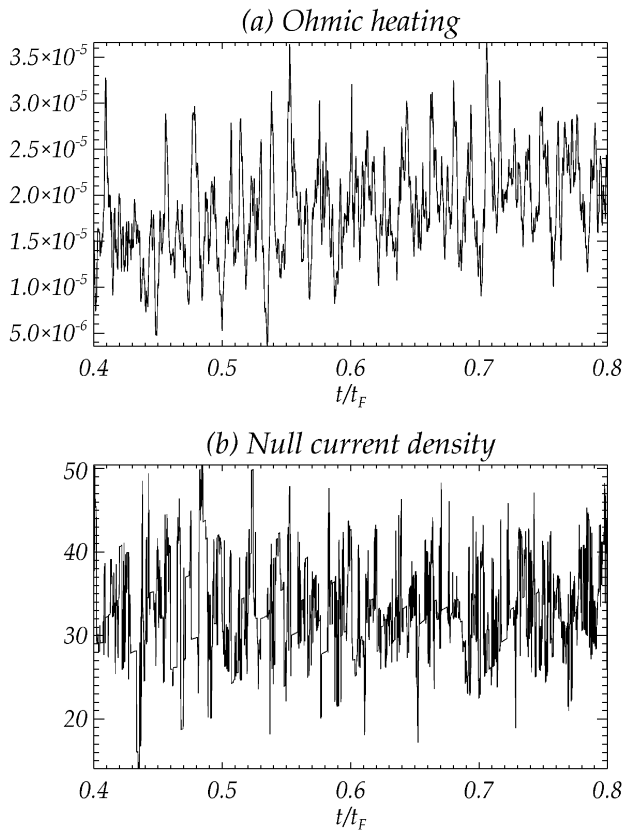}
  \caption{Time evolution of (a) ohmic heating and (b) the null current density, for the time interval $0.4<t/t_F<0.8$.}
  \label{fig:impulsive1}
\end{figure}

The energy plots in Fig. \ref{fig:energies1} show a further energy conversion after the initial transient phase (at $t\sim 0.02t_F$), which takes place in a completely new and different regime. This new phase is somewhat more chaotic than the first phase, as can be seen from the ohmic heating (Fig. \ref{fig:energies2}b) and the null current density (Fig. \ref{fig:energies2}c).

When the null current density has reached the critical value, $j_{crit}$, the resistivity switches off and reconnection stops. Nevertheless, the outward propagation of the magnetosonic waves, created by the initial rapid diffusion, leaves behind a reconnection flow (inwards at the top and bottom and outwards to the left and right) which drives a new regime of reconnection. This process evolves by constantly concentrating the current density at the null point during a short period of ideal evolution, and then quickly diffusing it, allowing a small amount of flux to be reconnected. The process repeats itself indefinitely in a random manner, reconnecting the field slowly, and in small sections. Hence, from this point, the null current density oscillates up and down, as the ohmic heating rises and falls at the same rate. This reconnection process is different from the first phase, as it is now driven by an actual reconnection flow and occurs in a quasi-steady manner. This type of reconnection is referred to as impulsive bursty reconnection \cite{priest86b}.

This impulsive bursty regime is a natural consequence of the reconnection-type forces left behind by the propagating waves, which build up the remaining current at the location of the null, combined with our choice of anomalous resistivity, which enables the possibility of switching the reconnection on and off in a continuous manner. Note, this impulsive bursty phase was not found in the higher plasma beta experiments \cite{Fuentes12}, and is also not found if a uniform instead of an anomalous resistivity is used. Hence, such a regime only occurs when (i) the microscopic properties of the plasma are favorable to induce an anomalous resistivity and (ii) the plasma pressure and the initial current accumulation are such that the return reconnection flows are strong enough to drive the initiation of the impulsive bursty reconnection.

\begin{figure}[t]
  \centering
  \includegraphics[scale=0.30]{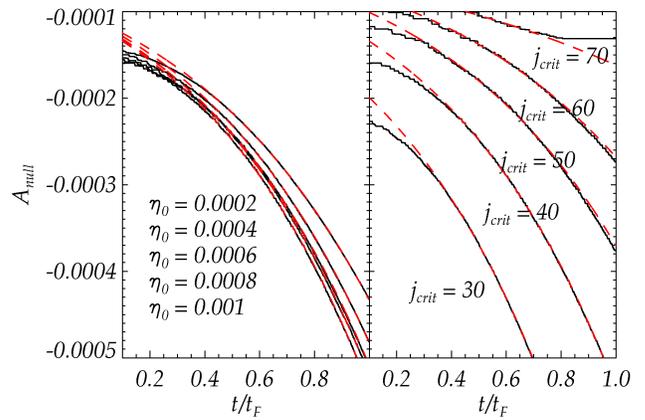}
  \caption{Plots of the null flux function, $A_{null}$, during the impulsive bursty regime, for different experiments varying the value of $\eta_0$ (left) and of $j_{crit}$ (right). All the curves are fitted by a function of the form $A_{null}=-\alpha{\rm e}^{\beta\,t/t_F}$ (dashed red), where both $\alpha$ and $\beta$ depend on the two parameters, $\eta_0$ and $j_{crit}$.}
  \label{fig:impulsive2}
\end{figure}

In Fig. \ref{fig:impulsive1}, we show the ohmic heating and the null current density during the impulsive bursty regime. The oscillation in both magnitudes is continuous, but does not show a well defined period. The application of a FFT (Fast Fourier Transform) indicates no dominant frequency, since the oscillations are stochastic. In future, a quantitative study of the impulsive bursty reconnection period is planned.

On the other hand, although the reconnection is not continuous, it always happens in the same direction, so that the overall change in flux function at the null is a monotonically decreasing function of time. The evolution of the null flux function is shown in Fig. \ref{fig:impulsive2}, for different values of both the resistivity, $\eta_0$, and the critical current, $j_{crit}$. In all cases, the flux function is modelled with a function of the form
\begin{equation}
A_{null}=-\alpha(\eta_0,j_{crit})\,{\rm e}^{\beta(\eta_0,j_{crit})\,t/t_F}\;.
\end{equation}
Experimentally, it is found that both $\alpha$ and $\beta$ increase monotonically with $\eta_0$ and decrease monotonically with $j_{crit}$.

Overall during this impulsive bursty phase, although the reconnection happens slowly at first, the long duration of this phase means more magnetic energy is converted than during the initial transient reconnection phase (as can be seen from Fig. \ref{fig:energies1}). An obvious question that arises is, whether this reconnection diffuses all the current at the null point in a finite amount of time. The answer is not obvious and needs to be investigated further.


\section{Discussion} \label{sec:sec5}


\subsection{Comparison with the high-beta scenario}

The experiment undertaken in Sec. \ref{sec:sec4} follows on from the study of \citet{Fuentes12}. The main difference between the two experiments is the background plasma pressure in which the non-resistive relaxation of the magnetic X-point evolution towards the formation of a current layer occurs. The initial state for the reconnection experiments in both cases is a current layer in a non-force-free equilibrium, with current extended along the separatrices, held by a non-zero plasma pressure gradient which is positive inside the cusps and negative outside the cusps. Here, we compare the {\it low-beta} (this paper) and the {\it high-beta} experiments \cite{Fuentes12}.

The key differences between the initial states of the two experiments are the dimensions and amplitude of the central current layer, which in the low-beta scenario is much thinner and more elongated with a peak of about 60 times larger than in the low beta case. Also, the fast and slow magnetosonic speeds in the low-beta case have different local values, while they have effectively the same speed in the high-beta case, and hence, are indistinguishable.

These aspects lead to a different structure and evolution for the pulses propagating out from the central diffusion region. The main similarities and differences in the results of the two experiments are as follows:\\
1) In both cases, the converted kinetic energy is much smaller (of the order of $1\%$) than the converted internal energy. Hence, the magnetic energy mainly causes ohmic heating of the plasma.\\
2) In both cases, there are initially four main propagating pulses. Two planar pulses come from the top and bottom of the diffusion region, carrying a deficit in plasma pressure and an inwards velocity flow. Two circular pulses come from the left and right vertices, carrying an excess in plasma pressure and an outwards velocity flow.\\
3) In both cases, the magnetic field perturbation (i.e. the current perturbation) shows a dominating $m=4$ mode, with a leading positive current front for the vertical and horizontal propagation across the field lines, and a leading negative current front for the diagonal propagation along the field lines. These structures create (or are created by) the required forces for a typical reconnection flow (see Fig. 8 in \citet{Fuentes12}).\\
4) In the low-beta case, the main conversion of energy occurs in the impulsive bursty phase rather than the initial transient phase. The impulsive bursty regime did not take place in the high-beta scenario, because the forces created by the reconnection flow were too small.\\
5) In the low-beta case, the amplitude of the outgoing magnetic perturbations is about $10^5$ times larger than in the high-beta case.\\
6) In the low-beta case, the structure of the overall perturbation is highly asymmetric, and its geometry changes with time, while the overall structure of the high-beta case perturbation remains quasi-circular. This is due to two factors. In the low-beta case, i) the initial diffusion region is much more elongated, and ii) the speeds of propagation are different inside and outside the cusps. In contrast, in the high-beta case, the slow and fast magnetosonic speeds take roughly the same value everywhere.\\


\subsection{Final conclusions}

Together with \citet{Fuentes12}, we have presented numerical studies on the rapid dissipation of a non-force-free current sheet, a non-trivial equilibrium obtained after the MHD viscous relaxation of a perturbed magnetic X-point.

The first important result is that, regardless of the initial energy and plasma state, the converted magnetic energy via reconnection goes almost entirely to direct heating of the surrounding plasma. Only a small portion is converted into kinetic energy, accelerating a complex pattern of fast and slow magnetosonic waves outwards from the diffusion region. Moreover, the amplitude of these waves is such that their latter dissipation via viscous damping, far from the null, is not important for the overall heating of the plasma.

However, under the assumptions of an anomalous resistivity,  over a longer time, the reconnection in the initial transient phase is not the only contribution to the heating of the surrounding plasma. Subsequently, the system around the null point enters a reconnection regime in which the reconnection flows themselves accumulate the current at the null bringing it to values above the critical level (which marks the level above which the resistivity switches on), causing a subsequent reconnection and repeating the process indefinitely. This impulsive bursty reconnection occurs in a continuous chain of reconnection events, which are able to transfer a higher amount of energy to the plasma than the initial spontaneous phase. However, this reconnection is stochastic and, due to numerical constraints, a quantitative parameter study is not possible with our present set up. The impulsive bursty reconnection regime described in this paper requires an anomalous resistivity, together with strong enough magnetic and plasma perturbations, such that the remaining reconnective flows can drive its initiation. Under this assumprions, in some realistic low-beta scenarios such as in the solar corona, this may be a plausible mechanism for the continuous release of magnetic energy.

A small amount of the energy released in the transient reconnection phase is transferred into kinetic energy, generating waves that travel out from the null at the local fast and slow speeds. These waves propagate with high-order modes, showing a complex family of fast and slow magnetosonic pulses, visible both in the plasma pressure and in the magnetic field. In a low plasma beta environment surrounding a magnetic null point, the magnetic field (current density) perturbation is larger than the plasma perturbation. Also, if the initial current layer is long enough, the two main planar pulses travelling in the direction perpendicular to the current layer become the dominant contribution and carry most of the kinetic energy released during the reconnection.

In a non-force-free reconnection scenario caused by the sudden onset of an anomalous resistivity, magnetic reconnection may be a contribution to heating the local plasma. This plasma heating occurs directly by ohmic heating first during a transient spontaneous reconnection phase and then, to a greater extent, in the long term release during an impulsive bursty reconnection phase. The continual reconnection at the null is similar to that of \citet{Longcope07}, who obtained a persistent electric field at the X-point, which continues flux transfer without accomplishing a significant energy dissipation. Nevertheless, our case is different in nature, mostly due to our anomalous resistivity, since it permits a further larger energy transfer than in the initial transient phase. 

The consequences of the present study, together with \citet{Fuentes12}, are of potential importance for flare and coronal heating, since it has been shown that the current layer formation in the solar atmosphere (due to, for example, motions of the photospheric footpoints of the magnetic field lines) can lead to spontaneous and driven null-point reconnection, as a source for direct plasma heating, in non-zero beta environments, plus a small contribution to accelerating waves out from the diffusion region, whose overall energetic consequences are insignificant (under the assumptions taken in this paper). Similar experiments for three-dimensional magnetic null point reconnection are to be carried out as a natural continuation of the present study.


\section*{Acknowledgements}

The authors would like to thank Prof. A.~W. Hood, Prof. D.~W. Longcope and Dr. I. De Moortel, for many useful discussions, good ideas and meaningful contributions to the present research. Computations were carried out on the UKMHD consortium cluster funded by STFC and SRIF. JFF is funded from the St Andrews Rolling Grant (ST/H001964/1).


\bibliographystyle{revtex4}


\end{document}